\documentstyle[aps,prd,twocolumn,psfig,floats,url]{revtex}
\def\Journal#1#2#3#4{{#1} {\bf #2}, #3 (#4)}

\def\PTP{Orig. Theor. Phys.}

\def\PLB{Phys. Lett. B}

\def\PRD{Phys. Rev. D}
\def\PTP{Prog. Theor. Phys.}
\def\ZETF{Zh. Eksp. Teor. Fiz.}

\def\etal{{\sl et al.}} 

\begin{document}
\preprint{\vbox{\hbox{UCB-PTH/XXXX}},
  \vbox{LBNL-XXXX},}
\wideabs{
\title{Energy Spectra of Reactor Neutrinos at KamLAND}
\author{Hitoshi Murayama, Aaron Pierce}

\address{
Center for Theoretical Physics, Department of Physics, 
University of California, 
Berkeley, CA~~94720, USA\\ 
Theory Group, 
Lawrence Berkeley National Laboratory, 
Berkeley, CA~~94720, USA \\
}

\date{\today}
\maketitle

\begin{abstract}
 The upcoming reactor neutrino experiment, KamLAND, has the ability 
to explore the Large Mixing Angle (LMA) solution to the solar 
neutrino problem.  Here, we investigate the precision to which 
KamLAND should be able to measure these parameters, utilizing the 
distortion of the energy spectrum of reactor neutrinos.  Incomplete 
knowledge of the fuel composition of the reactors will lead to 
some error on this measurement.  We estimate the size of this effect.
\end{abstract}
}
\narrowtext
\setcounter{footnote}{0}
\setcounter{page}{1}
\setcounter{section}{0}
\setcounter{subsection}{0} 
\setcounter{subsubsection}{0}


\vskip 0.3in

\section{Introduction}
For a long time we have been aware that there is a deficit of 
neutrinos emanating from the sun.  An apparent solution to this 
problem is the phenomenon of neutrino oscillations\cite{nuosc}.  
Particularly in light of the compelling atmospheric neutrino data 
from SuperKamiokande\cite{SuperKAtm}, we fully expect solar 
neutrinos to oscillate.  Oscillations between two flavors can be 
described effectively by two parameters: the mass difference,  
$\Delta m^2$, and a mixing angle, $\tan^2 \theta$.  There are four 
regions in the $\Delta m^2$, $\tan^2 \theta$ plane that explain the 
observed data on neutrinos from the sun.  KamLAND is designed to 
explore one of these regions, the Large Mixing Angle (LMA) solution, 
through the detection of reactor neutrinos.

KamLAND \cite{USKamLAND} is an experiment to be located at the old 
Kamiokande site in the Kamioka mine in Japan.  This location is of key 
importance to this experiment, as it is situated in the vicinity of 16 
nuclear power plants which will contribute a significant neutrino 
flux.  KamLAND consists of approximately 1 kiloton of liquid 
scintillator that will detect reactor neutrinos through the reaction:

\begin{equation}
\label{eqn:nureac}
p + \overline{\nu}_{e} \rightarrow n + e^{+}.
\end{equation}
The positron is then detected when it scintillates and when it 
annihilates an electron.  This annihilation, in delayed coincidence 
with the $\gamma$-ray from neutron capture, represents an easily 
recognizable signal.

In this letter, we assume that the solution to the solar neutrino 
problem is in the LMA region.  We explore the accuracy to which the 
KamLAND experiment can utilize the reaction of Eq.~(\ref{eqn:nureac}) 
to measure the parameters of a LMA solution.  In section 
\ref{sec:procedure} we review the basic procedure for computing the 
spectrum of neutrinos created at reactors.  In section 
\ref{sec:nevents}, we review how to compute the expected number of 
events at KamLAND.  In \ref{sec:results}, we describe the results of 
an analysis of the energy spectrum of the detected neutrinos.  Next, 
in section \ref{sec:spectrum} we address the question of systematic 
errors associated with an incomplete knowledge of the incident 
neutrino spectrum.  Finally, we conclude with a brief discussion of 
other possible systematics to be investigated.  We also briefly 
mention some of the implications of this measurement for future 
neutrino experiments.

\section{Determination of Energy Spectrum} \label{sec:procedure}

Because the mixing angle of $\nu_{e}$ in the mass gap responsible for 
the atmospheric neutrino oscillations is constrained to be small from 
CHOOZ reactor neutrino experiment \cite{CHOOZ}, the solar and atmospheric oscillations decouple to a very good approximation.  In 
this approximation, the important equation for the analysis of reactor 
neutrino oscillation becomes:
\begin{equation}  
\label{eqn:osc}
P(\overline{\nu}_{e} \rightarrow \overline{\nu}_{e})=1-\sin^2 2\theta 
\, \sin^2 \Bigl(\frac{1.27 \Delta m^2 (\mbox{eV$^2$}) L 
(\mbox{km})}{E (\mbox{GeV})}\Bigr).
\end{equation}
One of the advantages of the KamLAND design is that it is expected to 
have good energy resolution.  An energy resolution of 
$\frac{\sigma(E)}{E}=\frac{10\%}{\sqrt{E}}$ or better is anticipated, 
where $E$ is measured in MeV \cite{USKamLAND}.  As a result, if there 
are oscillations, one might hope to utilize the $E$ dependence of 
Eq.~(\ref{eqn:osc}) to assist in making an accurate 
measurement of the oscillation parameters.

Of course, in order to take advantage of this energy dependence, one 
must have knowledge of the energy dependence of the incident 
neutrinos.  That is to say, one is required to have a good knowledge 
of the unoscillated spectrum.  We discuss the determination of this 
spectrum here.

A number of short baseline experiments 
\cite{reactorspectra} have measured the energy spectrum of reactors 
at distances where oscillatory effects should be completely 
negligible.  A phenomenological parameterization of these spectra 
exists \cite{VogelSpectrum}, which depends on the isotope involved. 
Namely, Vogel and Engel find that:
\begin{equation}
\frac{dN_{\nu}}{dE_{\nu}}=e^{a_{0}+a_{1} E_{\nu}+ a_{2} E_{\nu}},
\end{equation}
where the fitted values of the  parameters $a_{i}$ are reproduced 
from \cite{VogelSpectrum} in Table~\ref{table:fitparam}.  This spectrum is given in 
units of $\frac{\overline{\nu}_{e}}{\rm MeV-fission}$.  Therefore, given 
this spectrum, it remains to determine how many fissions of each 
isotope there are. For a given reactor site, this will depend on 
three factors:
\begin{enumerate}
\item{ The thermal power of that reactor.}
\item{ The isotopic composition of the reactor fuel.}
\item{ The amount of thermal power emitted during the fissioning of a 
nucleus of a given isotope.}
\end{enumerate}.

\begin{table}
\caption{Parameters for $\frac{dN_{\nu}}{dE_{\nu}}$ parameterization. 
The resulting spectrum is given in units of 
$\overline{\nu}_{e}$/MeV-fission.}
\label{table:fitparam}
\begin{tabular}{|c|c|c|c|c|} \hline 
Isotope &  $^{235}$U & $^{239}$Pu & $^{238}$U & $^{241}$Pu \\ \hline 
$a_{0}$ &  0.870  & 0.896 & 0.976 & 0.793  \\ \hline 
$a_{1}$ &  -0.160 & -0.239 & -0.162 & -0.080 \\ \hline 
$a_{2}$ & -0.0910 & -0.0981 & -0.0790 & -0.1085 \\ \hline 
\end{tabular}
\end{table}

\begin{figure}[hbtp]
\psfig{file=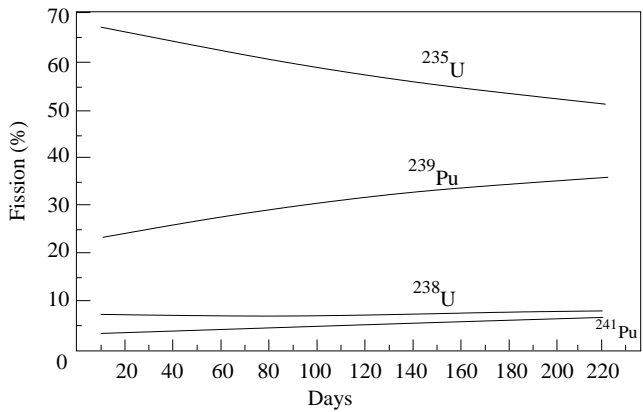,width=3.4in,angle=270}
\caption{The time dependent composition of reactor fuel. Taken from 
reference 
\protect\cite{Zacek}.} 
\label{fig:timecomp}
\end{figure}

For convenience, we reproduce the relevant reactor characteristics in 
Table~\ref{table:reactorparam}.  The maximum thermal power is given in 
the table.  This addresses the first point above.  Isotopic 
composition is a somewhat complicated issue, and we will investigate 
it more fully in section \ref{sec:spectrum}.  For the present, let us 
simply note that, in general, the time composition of the fuel varies 
roughly as in Fig.~\ref{fig:timecomp}.  For the analysis of section 
\ref{sec:results}, we take all reactors as having a composition 
varying as in this figure.  We assume that at the end of the 7 month 
cycle shown, the composition reverts back initial levels (refueling).  
Finally, for point three, we note that the thermal energy, 
$\epsilon_{i}$ associated with the fissioning of a given nucleon is 
known \cite{BoehmVogel}.  The relevant values are displayed in 
Table~\ref{table:energyperfission}.  With these pieces of information, 
we are able to determine the neutrino spectrum emitted from each 
reactor, $S(i,t,E)$.  Here, $i$ labels the reactor.

\begin{table}
\caption{Reactor parameters. Reproduced from reference 
\protect\cite{USKamLAND}. }
\label{table:reactorparam}
\begin{tabular} {l|c|c} \hline
Reactor Site & Distance (km) & Max. Thermal Power (GW)  \\ \hline
Kashiwazaki & 160 & 24.6 \\ \hline
Ohi & 180 & 13.7 \\ \hline
Takahama & 191 & 10.2 \\ \hline
Hamaoka & 213 & 10.6 \\ \hline
Tsuruga & 139 & 4.5 \\ \hline
Shiga & 81 & 1.6 \\ \hline
Mihama & 145 & 4.9 \\ \hline
Fukushima-1 & 344 & 14.2 \\ \hline
Fukushima-2 & 344 & 13.2 \\ \hline
Tokai-II & 295 & 3.3  \\ \hline
Shimane & 414 & 3.8 \\ \hline
Ikata &  561 & 6.0 \\ \hline
Genkai & 755 & 6.7 \\ \hline
Onagawa & 430 & 4.1 \\ \hline
Tomari & 784 & 3.3 \\ \hline
Sendai & 824 & 5.3 \\  \hline
\end{tabular} 
\end{table}

\begin{table}
\caption{Energy per fission of isotopes that make significant 
contributions to the thermal power of a reactor.}
\label{table:energyperfission}
\begin{tabular}{l|l|l|l|l}\hline
Fissioning Isotope &  $^{235}$U & $^{239}$Pu & $^{238}$U & $^{241}$Pu 
\\ \hline
Energy Per Isotope (MeV) & 201.7 & 205.0 & 210.0 & 212.4 \\ \hline
\end{tabular}
\end{table}

\section{Determination of Expected Number of Events} 
\label{sec:nevents}

Now that we have the initial spectrum in hand, we review how one 
finds the expected number of events at KamLAND.  To determine the 
number of neutrinos detected at KamLAND, one must convolve the cross 
section, $\sigma(E_{v})$, for the reaction shown in Eq.~(\ref{eqn:nureac}) 
with the reactor spectra, $S(i,t,E)$.  To lowest  
order the cross section is \cite{LowestXSection}: 
\begin{equation}
\sigma(E_{\nu})=\frac{2 \pi^2}{m_e^5 f \tau_{n}}p_{e} E_{e}.
\end{equation}
Here $f=1.69$ is the integrated Fermi function for neutron 
$\beta$-decay, $m_{e}$ is the mass of the electron, $E_{e}$ is the 
electron energy, and $p_{e}$ is the electron momentum.  The energy of 
the electron, to lowest order, is given by \cite{LowestXSection}: 
\begin{equation}
\label{eqn:Xsection}
E_{e}=E_{\nu}-1.293 \; \mbox{MeV}. 
\end{equation}
That is to say, the energy of the electron is basically the energy of 
the incident neutrino minus the proton-neutron mass difference.  In 
our numerical calculations, we used the cross section which takes 
into account the nucleon recoil, which may be found in 
\cite{FullXSection}.

The expected number of events at KamLAND is given by:
\begin{eqnarray}
\lefteqn{N(t, E_{\nu}, \Delta m^2, \sin^2 2 \theta)} \nonumber \\
&&=\sum_{i} 
\frac{S(i,t,E_{\nu})}{4 \pi d_{i}^2} \sigma(E_{\nu}) P(\Delta m^2, 
\sin ^2 2\theta).
\end{eqnarray}
Here, $P$ is the probability from Eq.~(\ref{eqn:osc}), 
$\sigma(E_{\nu})$ is the full cross section analogous to 
Eq.~(\ref{eqn:Xsection}), and $S(i,t,E_{\nu})$ is the initial energy  
spectrum, to be calculated as described above.  The values of the 
distances can also be found in Table~\ref{table:reactorparam}.

\section{Results} \label{sec:results}

\begin{figure}[hbtp]
\psfig{file=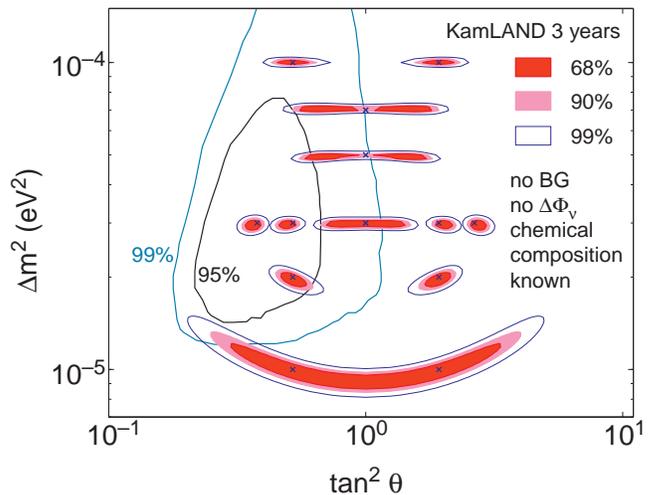,width=3.4in}
\caption{The expected measurement of $\Delta m^2$ and $\tan^2 
\theta$. Contours for 68\%, 90\% and 99\% CL are shown.The LMA solution 
to the solar neutrino problem is overlayed 
\protect\cite{Gonzales}.}
\label{fig:contours}
\end{figure}

With the expected number of events in hand, it is only a matter of a 
simple $\chi^2$ analysis to fit for the oscillation parameters.  In 
this section, we ignore background effects.  To estimate the precision 
to which KamLAND could measure the oscillation parameters, we assumed 
a 3 kt-year exposure where all reactors operated at 78\% of their 
maximum capacity.  It was assumed that the fuel composition of each 
reactor varied as shown in Fig.~\ref{fig:timecomp}.  Perfect detector 
efficiency was assumed.  To lowest order, this is not a bad 
assumption, given the recognizable delayed coincidence signal.  The 
measurements that KamLAND could be expected to perform with these 
assumptions are shown in Fig.~\ref{fig:contours}.  Contours were 
generated by finding the minimum $\chi^2$, and then calculating the 
confidence levels for two degrees of freedom ($\tan^2 \theta$ and 
$\Delta m^2$).  Data were binned in $0.5$ MeV bins.  The fit was done 
for visible energies\footnote{Here, visible energy is defined as the 
energy seen in the detector from the $e^{+}$ and its annihilation, 
$E_{\it visible}=E_{e^{+}}+m_{e^{-}}$.} between $E_{\it visible}=1.22$ 
MeV and $E_{\it visible}=7.0$ MeV. The first bin is smaller, since our 
parameterization of the incident spectra is only good above 
$E_{\nu}=2$ MeV. The cutoff at high energies is to avoid the low 
statistics bins, where there is not much statistical discrimination, 
and Poisson statistics would be necessary.

Since the oscillations in this case depend on the mixing angle only 
through $\sin^2 2\theta$, there is a two-fold degeneracy in the 
measurement (hence the reflection symmetry about $\tan^2 \theta=1$).  
However, the LMA solution, which is overlayed, does not posses such a 
symmetry, so it is necessary to plot against $\tan^2 \theta$ and not 
$\sin^2 2\theta$ \cite{darkside}.  

\begin{figure}[hbtp]
\psfig{file=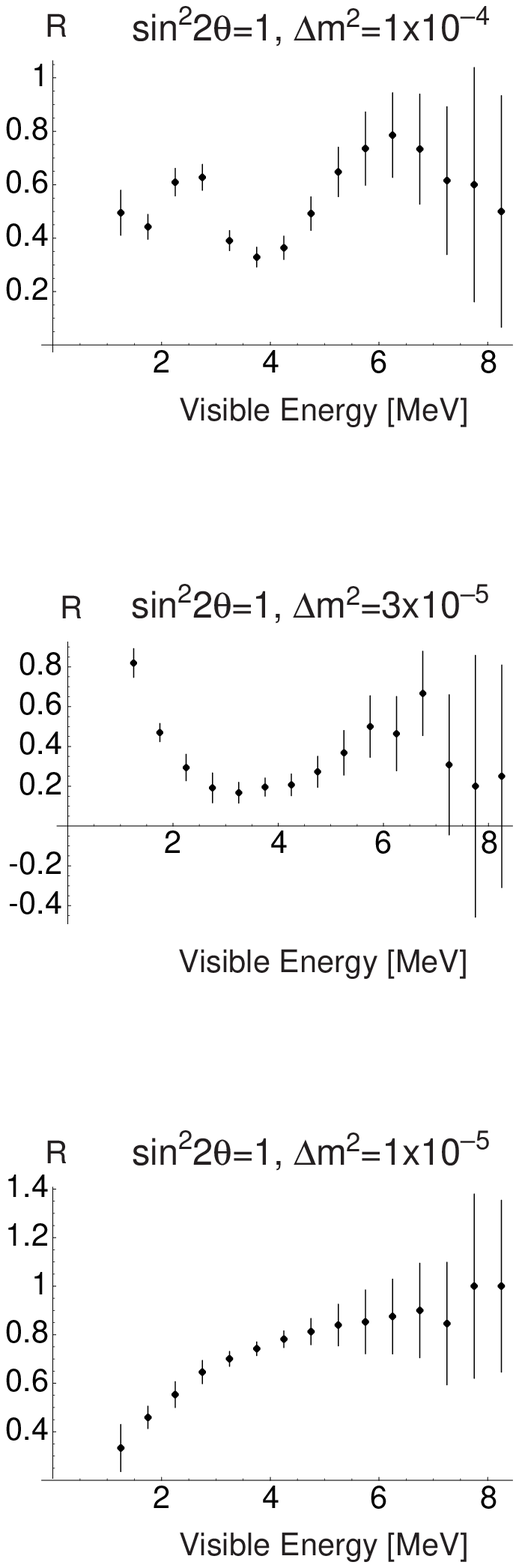,height=8.5in}
\caption{The ratio, R, of the number of events seen, to the number of 
events expected if there were no oscillations.  The error bars are 
statistical.}
\label{fig:Rplot}
\end{figure}

From the figure, it is clear that KamLAND is able to make a very 
accurate measurement of the $\Delta m^2$ parameter in particular.  To 
see why this is so, it is instructive to look at the quantity:

\begin{equation}
R=\frac{\# \mbox{Observed}}{\# \mbox{No Oscillation}}.
\end{equation}
Plots of this quantity are shown in Fig.~\ref{fig:Rplot}.  This should 
essentially be the oscillation probability.  By determining the 
position of the dip in the oscillation, one is able to determine the 
value of $\Delta m^2$.  From the figure, one can see that the location 
of the dip can be well determined by KamLAND if the oscillations 
parameters lie in the LMA solution to the solar neutrino problem.
 
\section{Fuel Composition Effects} \label{sec:spectrum}

Let us now address the question of the effect of the fuel 
composition.  To calculate the evolution of the different components 
of reactor fuel is a complicated business.  KamLAND expects to obtain 
the time dependent compositions from the power companies involved.  
This section serves to demonstrate that this is an important piece of 
information for this experiment. 

To estimate the effect that an incomplete knowledge of the fuel 
composition would have on the measurement, we took the extreme cases.  
We look at the neutrino spectrum that would result from taking the 
composition to be as illustrated in Fig.~\ref{fig:contours}, we refer 
to this as the ``true composition.''  We then tried to fit the 
observed data that were generated from the true composition with the 
expected number of events from three different compositions:
\begin{enumerate}
\item  The true composition, 
\item A composition that stays constant and equal to the composition 
at $t=0$,
\item A composition that stays constant and equal to the composition 
at $t=220$ days.
\end{enumerate}

\begin{figure}[hbtp]
\psfig{file=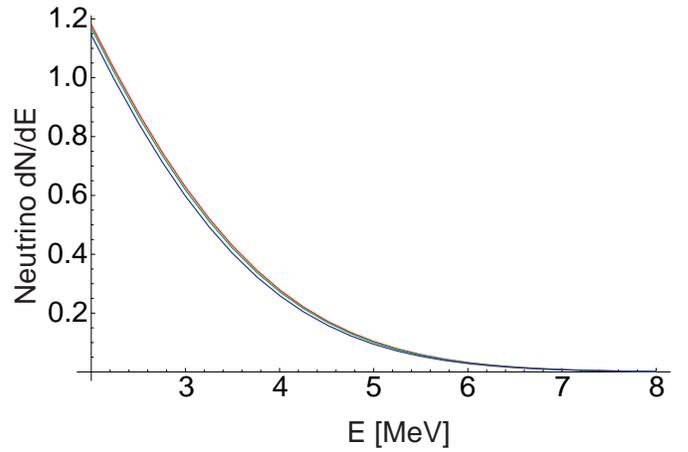,width=3.4in}
\caption{Neutrino spectra generate using the initial, final and 
average fuel composition.Although the differences here are somewhat 
difficult to discern, they are move easily visible in 
Fig.~\ref{fig:difference}.}
\label{fig:spectra}
\end{figure}

\begin{figure}[hbtp]
\psfig{file=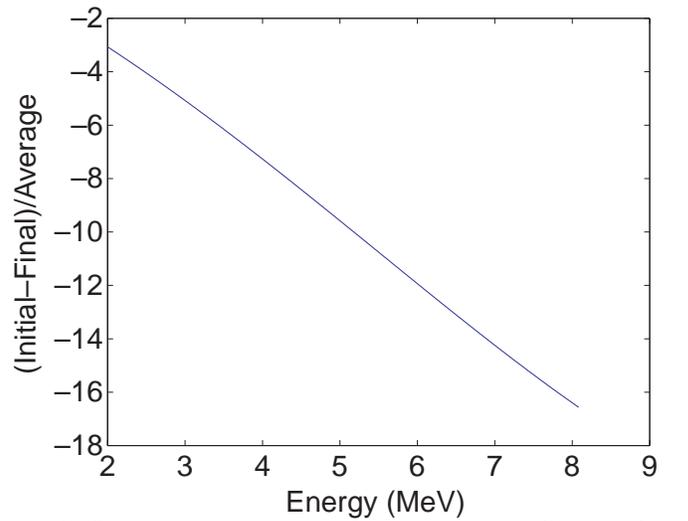,width=3.4in}
\caption{Here we have plotted the percentage difference in the 
initial and final spectra.  More precisely, we have plotted the 100 times 
the difference between the initial and final spectra, divided by the 
average spectrum.}
\label{fig:difference}
\end{figure}

The different spectra are shown in Fig.~\ref{fig:spectra}.  Although 
the basic shapes are relatively consistent, there are important 
differences, as can be seen by examining Fig.~\ref{fig:difference}, 
where we plot the difference between the spectra, normalized to the 
average spectrum.  At the higher energies, differences in the initial 
and final spectra can reach over 15\%.

\begin{figure}[hbtp]
\psfig{file=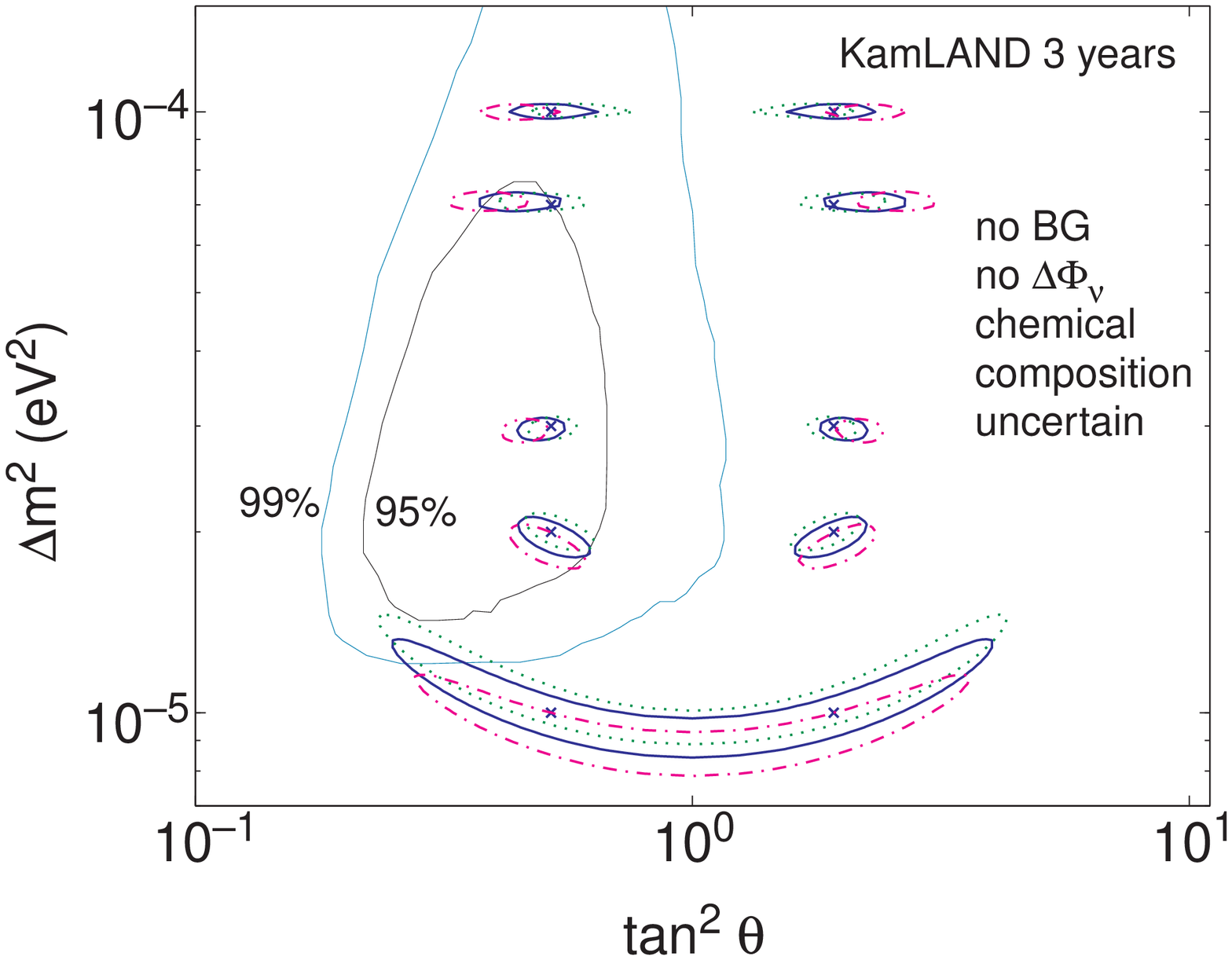,width=3.4in}
\caption{90 \% confidence level contours for measurements of $\Delta m^2$ 
and $\tan^2 \theta$.  The contours were generated by varying the 
assumptions about the composition of the fuel.  For further 
explanation, see the text.} 
\label{fig:comparespec.eps}
\end{figure}

The difference between the three fits done in this way should give an 
estimate of the systematic error involved.  A contour plot showing the 
measurements in the ($\tan^2 \theta$, $\Delta m^2$) plane is shown in 
Fig.~\ref{fig:comparespec.eps}.  Although the values of the measured 
quantities are not drastically affected, particularly in the cases of 
moderate $\Delta m^2$, fuel composition clearly is an important 
systematic error to be considered.  It is also of interest to see how 
incorrect assumptions about fuel composition degrades the fit.  We 
show the $\chi^2$ for 10 degrees of freedom in Table~\ref{table:chis}. In 
particular, the $\chi^2$ is not nearly as good for the fits with the 
final spectrum.

\begin{table}
\caption{The fits corresponding to three different fuel compositions. 
For each point in ($\Delta m^2$,$\tan^2 \theta$) space, the events 
incident on KamLAND were generated with a time dependent (T) 
composition.  Fits were done with the time dependent (T), initial 
(I), and final (F), composition.}
\label{table:chis}
\begin{tabular}{l|l|l|l|r|r}
Comp. & $\Delta m^2$ & $\tan^2 \theta$ & Fit $\Delta m^2$ & Fit 
$\tan^2 \theta$ & $\chi^2$ (10 dof) \\ \hline \hline
T       & $1 \times 10^{-5}$ & 0.52, 1.92  &  $9.5 \times 10^{-6}$& 0.64, 
1.55
        & 12.89 \\ \hline
I       & $1 \times 10^{-5}$ & 0.52, 1.92  &  $1.1 \times 10^{-5}$& 0.43, 
2.30            & 9.64  \\ \hline
F       & $1 \times 10^{-5}$ & 0.52, 1.92 & $8.7 \times 10^{-6}$& 0.80, 
1.24
        & 21.67  \\ \hline \hline
T       & $2 \times 10^{-5}$ & 0.52, 1.92 & $1.9 \times 10^{-5}$ & 0.54, 
1.87 
        & 12.42\\ \hline
I       & $2 \times 10^{-5}$ & 0.52, 1.92 & $ 2.0 \times 10^{-5}$ & 0.53, 
1.87            &  9.50\\ \hline
F       & $2 \times 10^{-5}$ & 0.52, 1.92 & $ 1.9 \times 10^{-5}$ & 0.50, 
2.01            & 18.26 \\ \hline \hline
T       & $3 \times 10^{-5}$ & 0.52, 1.92 & $3.0 \times 10^{-5}$ & 0.50, 
2.01
        & 12.17  \\ \hline
I       & $3 \times 10^{-5}$ & 0.52, 1.92 & $3.0 \times 10^{-5}$ & 0.51, 
1.94 
        & 10.2 \\ \hline
F       & $3 \times 10^{-5}$ & 0.52, 1.92 & $3.0 \times 10^{-5}$ & 0.46, 
2.16 
        & 16.19 \\ \hline \hline
T       & $7 \times 10^{-5}$ & 0.52, 1.92 & $7.2 \times 10^{-5}$ & 0.45, 
2.23 
        & 10.73 \\ \hline
I       & $7 \times 10^{-5}$ & 0.52, 1.92 & $7.0 \times 10^{-5}$ & 0.50, 
2.01 
        & 11.96 \\ \hline
F       & $7 \times 10^{-5}$ & 0.52, 1.92 & $7.2 \times 10^{-5}$ & 0.39, 
2.58
        & 11.57 \\ \hline \hline
T       & $1 \times 10^{-4}$ & 0.52, 1.92 & $1.0 \times 10^{-4}$ & 0.51, 
1.94 
        & 12.96 \\ \hline
I       & $1 \times 10^{-4}$ & 0.52, 1.92 & $1.0 \times 10^{-4}$ & 0.58, 
1.72
        & 8.58 \\ \hline
F       & $1 \times 10^{-4}$ & 0.52, 1.92 & $1.0 \times 10^{-4}$ & 0.45, 
2.23
        & 21.87
\end{tabular}
\end{table}

\section{Flux Uncertainty Effects}

\begin{figure}[hbtp]
\psfig{file=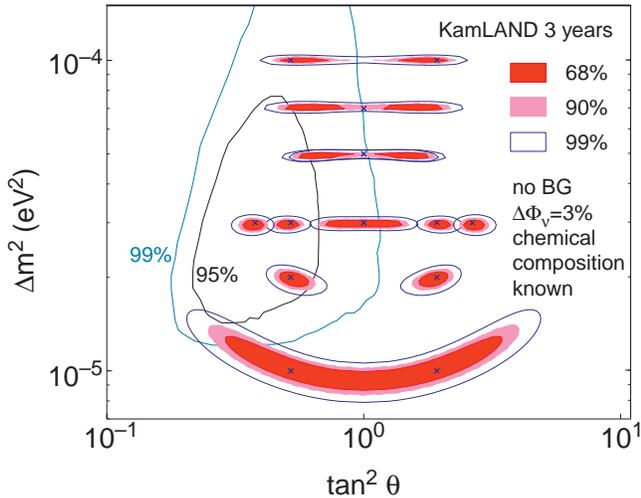,width=3.4in}
\caption{The contours that arise from including the flux uncertainty 
in the $\chi^2$.} 
\label{fig:flux}
\end{figure}

Finally, we relax our assumption that KamLAND will have complete 
knowledge of the overall flux normalization.  We assume that the flux 
is known with a three percent error, consistent with the size of the 
errors in the $\beta$-spectroscopy experiment at the G\"{o}sgen 
reactor \cite{Zacek}.  We then can compute a $\chi^2$, that takes 
this uncertainty into account.  In particular, we write:
\begin{equation}
\label{eqn:modchi}
\chi^2={\rm min}\left(\chi^2_{bins}+\frac{(N-1.0)^2}{\sigma_N^2}\right).
\end{equation}
Here, $\sigma_N^2=.03^2$.  We vary the possible normalizations, $N$, 
and take the minimum of the right hand side.  Of course, the 
$\chi^2_{bins}$ is a function of $N$ as well, as the theoretical 
prediction depends on the incident flux.  Note, in general, this 
procedure will provide a smaller $\chi^2$, at each set of oscillation 
parameters, resulting in larger contours.  From Fig.~\ref{fig:flux}, we can see 
that the contours are enlarged slightly.  In particular, the 
uncertainty in the overall flux causes a slight degradation of the 
measurement of the mixing angle.  This is a well-known and important 
effect.  For completeness, we tabulate the values of $\chi^2$ in 
Table~\ref{table:fluxchis}.

\begin{table}
\caption{The fits corresponding to three different fuel compositions. 
For each point in ($\Delta m^2$,$\tan^2 \theta$) space, the events 
incident on KamLAND were generated with a time dependent 
composition.  Fits were done either allowing the overall flux to float 
(Float), or keeping it fixed at the predicted value. (Fixed).}
\label{table:fluxchis}
\begin{tabular}{l|l|l|l|r|r}
Comp. & $\Delta m^2$ & $\tan^2 \theta$ & Fit $\Delta m^2$ & Fit 
$\tan^2 \theta$ & $\chi^2$ (10 dof) \\ \hline \hline
Fix     & $1 \times 10^{-5}$ & 0.52, 1.92  &  $9.5 \times 10^{-6}$& 
0.56, 1.55
        & 12.89 \\ \hline
Float   & $1 \times 10^{-5}$ & 0.52, 1.92  &  $1.0 \times 10^{-5}$& 
0.56, 1.79      & 12.38  \\ \hline \hline
Fix     & $2 \times 10^{-5}$ & 0.52, 1.92 & $1.9 \times 10^{-5}$ & 
0.54, 1.87 
        & 12.42\\ \hline
Float   & $2 \times 10^{-5}$ & 0.52, 1.92 & $ 2.0 \times 10^{-5}$ & 
0.54, 1.87      &  12.06\\ \hline \hline
Fix     & $3 \times 10^{-5}$ & 0.52, 1.92 & $3.0 \times 10^{-5}$ & 
0.50, 2.01
        & 12.17  \\ \hline
Float   & $3 \times 10^{-5}$ & 0.52, 1.92 & $3.0 \times 10^{-5}$ & 
0.51, 1.94 
        & 11.88 \\ \hline \hline
Fix     & $7 \times 10^{-5}$ & 0.52, 1.92 & $7.2 \times 10^{-5}$ & 
0.45, 2.23 
        & 10.73 \\ \hline
Float   & $7 \times 10^{-5}$ & 0.52, 1.92 & $7.0 \times 10^{-5}$ & 
0.61, 1.63
        & 9.74 \\ \hline \hline
Fix     & $1 \times 10^{-4}$ & 0.52, 1.92 & $1.0 \times 10^{-4}$ & 
0.51, 1.94 
        & 12.96 \\ \hline
Float   & $1 \times 10^{-4}$ & 0.52, 1.92 & $1.0 \times 10^{-4}$ & 
0.58, 1.71
        & 11.71
\end{tabular}
\end{table}

\section{Experimental Systematics}

Although we are not in a position to make a quantitative study of the 
experimental systematics at KamLAND, a few brief comments are 
possible.  First of all, in the preceding analysis, we have neglected 
the contribution from the backgrounds.  In the reactor experiment, 
there is a distinctive delayed neutron-capture signature, which 
results in a substantial reduction of backgrounds.  The background 
level was conservatively estimated to be 20:1 \cite{USKamLAND} ({\it 
i.e.}\/, probably an overestimate).  It is not so unreasonable to 
ignore the backgrounds.  Moreover, one expects that backgrounds will 
be a relatively steeply falling function of energy.  So, if the energy 
where the interesting oscillation effects occur is high enough, 
then the assumption of no backgrounds is even safer.  This happens at 
the larger $\Delta m^2$ values, as can clearly be seen from Eqn.  (2).  
In addition, KamLAND hopes to get a handle on the backgrounds by 
utilizing the fact that power usage, and hence neutrino flux, varies 
seasonally \cite{USKamLAND}.  The amount of power produced over time 
will be available from the power companies.  

We would like to briefly 
mention other possible sources of systematic uncertainties.  As 
mentioned earlier in the paper, there is an issue of the energy resolution of the 
detector.  This could smear out the location of the ``dip'' that is so 
nicely seen, in Fig.~\ref{fig:Rplot}(a).  However, the resolution expected is 
better than the size of the bins and hence is not expected to affect 
the results.  A more difficult issue would be to accurately callibrate 
the energy measurement of the detector of this size.  There are plans 
to do so; see \cite{USKamLAND}.  There will be 
a source of backgrounds from geological neutrinos in the 2-3 
MeV region which may or may not be important.  However, a 
simple energy cut should remove them completely.  

\section{Conclusions}

The KamLAND reactor neutrino experiment should allow for an accurate 
measurement of $\Delta m^2$, as long as the solution to the solar 
neutrino problem is in the LMA region.  Incomplete knowledge of the 
fuel composition at the reactors will represent an important source of 
experimental error.  In this letter, we assumed perfect detector 
efficiency, and that backgrounds were well understood.  It is 
interesting to note that the precision to which KamLAND is able to 
measure $\Delta m^2$ may effect the ability of a muon source neutrino 
factory to extract information about the CP violating phase in the MNS 
matrix.  CP violation is an phenomenon that requires at least three 
generations, the effects of the sub-leading oscillation is crucial, 
and the matter effect needs to be accurately subtracted in order to 
determine $\delta_{CP}$.  The measurements at KamLAND should prove 
crucial for this purpose.

\section{Acknowledgements}

This was supported in part by the Director, Office of Science, Office 
of High Energy and Nuclear Physics, Division of High Energy Physics 
of the U.S. Department of Energy under  Contract DE-AC03-76SF00098 
and in part by the National Science Foundation  under grant 
PHY-95-14797.  AP is also supported by a National Science Foundation 
Graduate Fellowship.

\section*{Note Added}
During the final preparation of this manuscript, we became aware of
papers on a similar topic \cite{Barger,Barbieri}.  Both of them
discussed how accurately KamLAND will determine the oscillation
parameters.  Ref.~\cite{Barger} further included the three-generation
mixing effects, while Ref.~\cite{Barbieri} studied $\Delta m^2 > 2
\times 10^{-4}$~eV$^2$ as well.  Neither of them discusses the effect
of fuel composition and overall flux normalization quantitatively,
however.  Other conclusions appear to be consistent.


\begin{thebibliography}{99}
\bibitem{nuosc}
Z. Maki, M. Nakagawa, and S. Sakata, \Journal{\PTP}{28}{870}{1962}.
B. Pontecorvo. \Journal{\ZETF}{52}{1717}{1967}.
\bibitem{SuperKAtm}
Super-Kamiokande Collaboration (Y. Fukuda et al.), 
\Journal{\PLB}{B436}{33}{1998}.
\bibitem{USKamLAND} 
``Proposal for US Participation in KamLAND,'' March 1999, 
\url{http://kamland.lbl.gov/KamLAND.US.Proposal.pdf}
\bibitem{CHOOZ} M.~Apollonio {\it et al.}
Phys.\ Lett.\  {\bf B466}, 415 (1999)
[hep-ex/9907037].
\bibitem{reactorspectra}
A.A. Hahn \etal \Journal{\PLB}{218}{365}{1989}.
K. Schreckenbach \etal \Journal{\PLB}{160}{325}{1985}.
\bibitem{VogelSpectrum}
P. Vogel and J. Engel. \Journal{\PRD}{39}{3378}{1989}.
\bibitem{BoehmVogel}
F. Boehm and P.Vogel. {\underline The Physics of Massive Neutrinos.} 
(New York: Cambridge University Press) 1992.
\bibitem{Gonzales}
LMA contour taken from C. Gonzalez-Garcia at {\tt 
http://ific.uv.es/{}$^\sim$penya/2nu.html}.
\bibitem{Zacek}
G. Zacek, et al., \Journal{\PRD}{34}{2621}{1986}.
\bibitem{LowestXSection}
P. Vogel. \Journal{\PRD}{29}{1918}{1984}.
\bibitem{FullXSection}
P. Vogel. and J.F. Beacom. \Journal{\PRD}{60}{053003}{1999}.
\bibitem{darkside} A.~de Gouvea, A.~Friedland and H.~Murayama.
Phys.\ Lett.\  {\bf B490}, 125 (2000)
[hep-ph/0002064].
\bibitem{Barger}
V. Barger, D. Marfatia, and B. Wood. {\tt hep-ph/0011251}.
\bibitem{Barbieri}
R.~Barbieri and A.~Strumia,
{\tt hep-ph/0011307}.
\end{thebibliography}
\end{document}